\documentclass[prc,aps,showpacs]{revtex4}
\usepackage{graphicx}

\begin{document}

\title{\bf Shell-model calculations of two-neutrino double-beta decay rates of $^{48}$Ca  
with GXPF1A interaction} 

\author{M. Horoi$^a$, S. Stoica$^b$ and B. A. Brown$^c$}

\affiliation{
$^{a}$Department of Physics, Central Michigan University, Mount Pleasant,
Michigan 48859, USA \\
$^{b}$Horia Hulubei National Institute for Physics and 
Nuclear Engineering,
 P.O. Box MG-6, 077125 Magurele-Bucharest, Romania\\
$^{c}$National Superconducting Cyclotron Laboratory 
and Department of Physics and Astronomy,
 Michigan State University, East Lansing, Michigan 48824, USA} 

\vskip.5cm

\begin{abstract} 
 The two-neutrino double beta decay matrix elements and half-lives of $^{48}$Ca, are 
calculated within a shell-model approach for transitions to the ground state and to 
the  $2^+$ first excited state of $^{48}$Ti. We use the full $pf$ model space and 
the GXPF1A interaction,
which was recently proposed to describe the spectroscopic properties of the nuclei in
the nuclear mass region A=47-66.
Our results are $T_{1/2}(0^{+}\rightarrow 0^{+})$ = $3.3\times 10^{19}$ $yr$ and  
$T_{1/2}(0^{+}\rightarrow 2^{+})$ = $8.5\times 10^{23}$ $yr$. 
The result for the decay to the $^{48}$Ti 0$^+$ ground state is in 
good agreement with experiment. The half-life for the decay to the 2$^+$
state is two orders of magnitude larger than obtained previously.

\end{abstract}

\pacs{23.40.Bw, 21.60.Cs, 23.40.Hc}

\date{September 29, 2006}
\maketitle



At present, the double-beta ($\beta\beta$) decay is the 
most sensitive process for direct measurements of the electron neutrino mass.\cite{[SC98]}-\cite{[EE04]}
For deriving the neutrino mass one needs, on one hand, experimental half-lives 
for the neutrinoless $\beta\beta$ ($0\nu\beta\beta$) decay mode and, on the other hand, 
theoretical values of the nuclear matrix elements (NME) entering these half-lives formulae.   
 
After many years of intense investigations and debate on different nuclear 
structure methods, accurate calculation of the NME relevant for $\beta\beta$ decay 
remains a challenging issue. Since many 
$\beta\beta$ emitters are nuclei with open shells, the
proton-neutron random phase approximation (pnQRPA) and its extensions, have been 
the most used methods to perform such calculations.\cite{[VOG86]}-\cite{[STO]} 
However, due to the significant progress in shell-model (SM) configuration
mixing 
approaches, there are now calculations performed with these methods 
for several nuclei.\cite{[ZBR90]}-\cite{[NOW05]} In spite of their success in getting 
agreement with the experimental half-lives of the 
two-neutrino $\beta\beta$ ($2\nu\beta\beta$) decay mode, both 
pnQRPA- and SM-based approaches have some shortcomings that limit their predictive 
power for the NME in the case of the more interesting $0\nu\beta\beta$ mode. 
For example, within pnQRPA methods the NME exhibit a high sensitivity to 
the renormalization of the particle-particle strength in the $1^+$ 
channel, while within SM one has to severely truncate the 
model space in order to make the diagonalization procedure tractable. 
In order to better understand the source of uncertainties of the NME calculations 
for $\beta\beta$ decay, a systematic comparison between calculations performed with both 
types of methods is needed. This comparison will become more feasible 
as the computational power of the shell model methods expand to 
treat larger model spaces.
The effective two-body interaction employed 
is also important, since the $B(GT^+)$
strengths are especially sensitive to these interactions.

\begin{figure*}
\includegraphics[width=7cm,angle=-90]{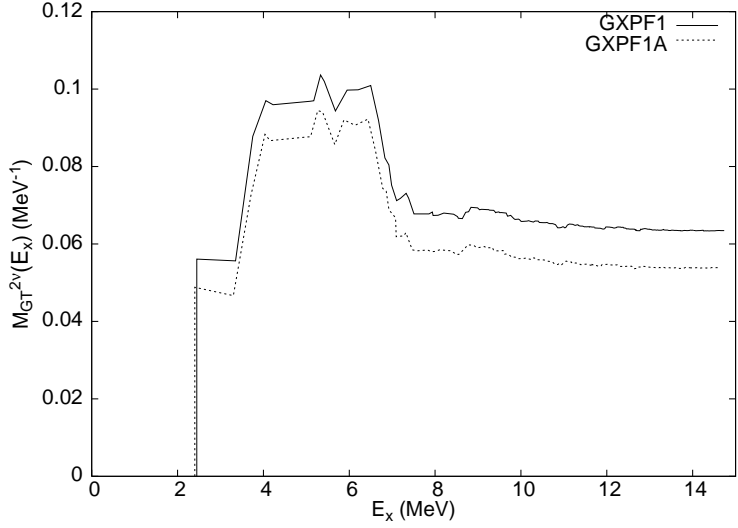}
\caption{\label{f1} The $2 \nu \beta\beta$ decay running matrix element of Eq. (1) as
a function of the excitation energy of the intermediate $1^{+}$ states in
$^{48}$Sc, for the GXPF1 and GXPF1A interactions.} 
\end{figure*}

SM calculations for $\beta\beta$ decay can now be 
carried out rather accurately for $^{48}$Ca. 
 Zhao, Brown and Richter \cite{[ZBR90]} calculated the $2\nu\beta\beta$ 
NME of $^{48}$Ca in a large basis SM space using the OXBASH code with the
 MH (Muto and Horie)
 \cite{[MH84]} and MSOBEP \cite{[RJB90]} two-body interactions. Their predicted $T^{2\nu}_{1/2}$ 
 is smaller than the experimental one. They also made an analysis 
of the distribution of the $B(GT^-)$, $B(GT^+)$ and $M^{2\nu}_{GT}$ components over 
the $1^+_k$ excitation 
energies in the intermediate nucleus ($^{48}$Sc), which helps 
better understand the quenching of 
the NME for the $2\nu\beta\beta$ decay mode. 
Caurier, Poves and Zuker \cite{[CPZ90]} performed 
a full $pf$ shell calculation of the NME for the $2\nu\beta\beta$ decay mode, 
both for the transitions 
to the g.s. and to the $2^+_1$ of $^{48}$Ti. 
Their calculations were carried out with the ANTOINE code.\cite{[ANT]} 
As an effective interaction they   
used the Kuo-Brown G-matrix \cite{[KB68]} with minimal 
monopole modifications, KB3.\cite{[PZ81]} We will discuss their results
together with our new results below.
 
In this paper we use the recently proposed GXPF1A two-body effective interaction, 
which has been successfully 
tested for the $pf$ shell \cite{[gx1]}-\cite{[ni56]}, to perform $2\nu\beta\beta$ decay 
calculations for $^{48}$Ca. 
Our goal is to obtain the values of the NME for this decay mode, both for transitions to the 
g.s. and to the $2^+_1$ state of $^{48}$Ti, with increased degree of confidence, 
which will allow us in the next 
future to address similar calculations for the $0\nu\beta\beta$ decay mode of 
this nucleus.\cite{reta95}  
The $2\nu\beta\beta$ transitions to excited states have longer half-lives, as compared
with the transitions to the g.s., due to the reduced values of the
corresponding phase spaces. 
Positive results for the $2\nu\beta\beta$ 
 decay of $^{100}$Mo, were recently reported.\cite{mo05}.


For the $2\nu\beta\beta$ decay mode the relevant NME are of Gamow-Teller type,  
 and have the following expressions \cite{[SC98]}- \cite{[EE04]}:

\begin{equation}
M^{2\nu}_{GT}(0^+) = \sum_k 
\frac{
\langle 0_f\vert\vert \sigma\tau^-\vert\vert 1^+_{k}
\rangle 
\langle 1^+_{k}\vert\vert \sigma\tau^-
\vert\vert 0_i\rangle
}{E_k + E_0}\ , 
\label{eq1}
\end{equation}

\noindent 
 for the g.s. to g.s. transition, and 

\begin{equation}
M^{2\nu}_{GT}(2^+) = \frac{1}{\sqrt{3}}
\sum_k \frac{\langle 2^+_f\vert\vert \sigma\tau^-\vert\vert 1^+_{k}
\rangle 
\langle 1^+_{k}\vert\vert \sigma\tau^-
\vert\vert 0_i\rangle}{(E_k + E_2)^3} 
\label{eq2}
\end{equation}

\noindent
for the g.s. to  $2^+_1$ transition. 
Here $E_k$ is the excitation energy of the $1^+_k$ state of $^{48}$Sc
and $E_0 = \frac{1}{2}Q_{\beta\beta}(0^+) + \Delta M$,  
$E_2 = \frac{1}{2}Q_{\beta\beta}(2^+) + \Delta M$.
$Q_{\beta\beta}(0^+)$ and $Q_{\beta\beta}(2^+)$ are the Q-values 
corresponding to the $\beta\beta$ 
decays to the g.s. and the $2^+_1$ excited state of the parent nucleus 
($^{48}$Ti) and $\Delta M$ is the 
$^{48}$Ca -$^{48}$Sc mass difference.

The $\beta\beta$ half-live expression is given by

\begin{equation}
\left[T^{2\nu,J}_{1/2}\right]^{-1} = 
F^{2\nu}_J\vert M^{2\nu}_{GT}(J)\vert^2 
\label{hlive}
\end{equation}

\noindent
where $F^{2\nu}_J$ are the phase space factors\cite{[SC98]}: 
$1.044\times10^{-17}\ yr^{-1}\ $MeV$^2$, 
corresponding to the g.s. to g.s. transition (J=0)
and $1.958\times10^{-19}\ yr^{-1}\ $MeV$^6$, corresponding 
to g.s. to $2^+_1$ transitions (J=2), respectively.  

\begin{figure*}
\includegraphics[width=7cm,angle=-90]{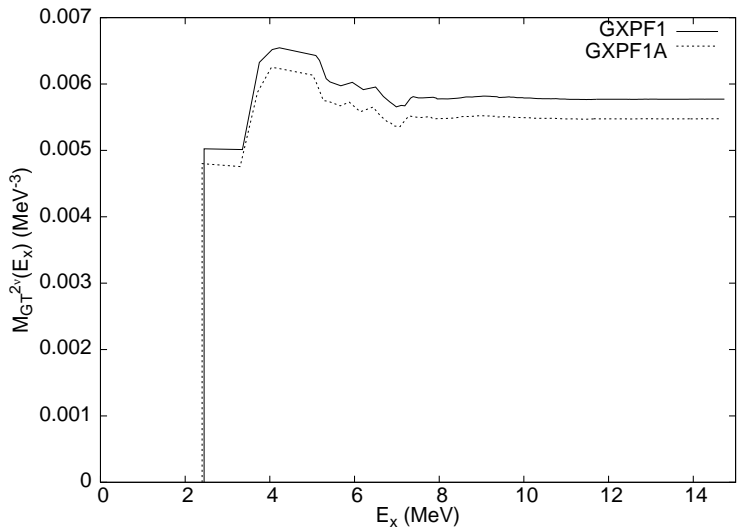}
\caption{\label{f2} The $2 \nu \beta\beta$ decay running matrix element of Eq. (2)
(up to a factor of $\sqrt{5}$)  as
a function of the excitation energy of the intermediate $1^{+}$ states in
$^{48}$Sc, for the GXPF1 and GXPF1A interactions.} 
\end{figure*}




The calculations were carried out in the full $pf$ model space using
the CMISHSM shell model-code\cite{[cmichsm]} and
 the GXPF1A interaction.
The most recent effective Hamiltonians, GXPF1 \cite{[gx1]}-\cite{[gx1ap]}
and GXPF1A \cite{[gx1a]} are derived from a microscopic calculation
by Hjorth-Jensen based on renormalized G matrix theory with the Bonn-C interaction
\cite{[rg]}, and are refined by a systematic fitting of the important
linear combinations of two-body matrix elements to low-lying states
in nuclei from A=47 to A=66. GXFP1A addresses some shortcoming of the 
GXPF1 interaction for the region of the neutron rich Sc, Ti and Ca 
isotopes\cite{[gx1a]} that is relevant for this study.
An advantage of using the full $pf$ model space is that the Ikeda sum rule
is exactly satisfied.

\begin{figure*}
\includegraphics[width=7cm,angle=-90]{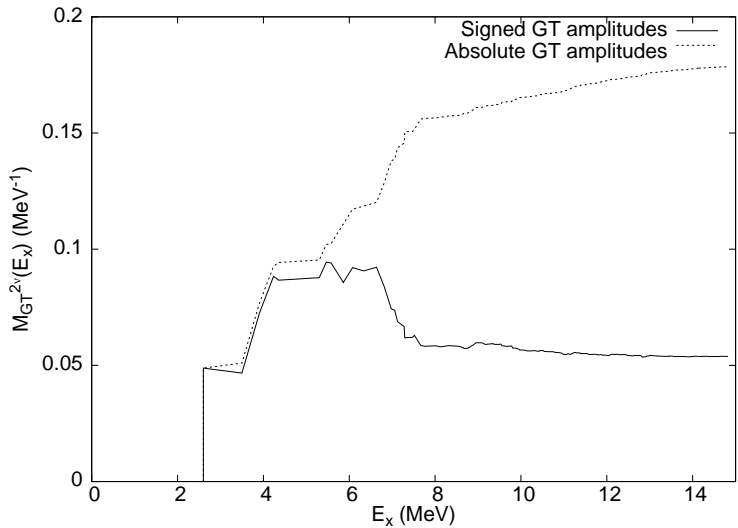}
\caption{\label{f3} The $2 \nu \beta\beta$ decay running matrix element of Eq. (1) as
a function of the excitation energy of the intermediate $1^{+}$ states in
$^{48}$Sc compared with the similar sum where the absolute values of the GT 
matrix elements are used.} 
\end{figure*}

In the calculation of the NME, Eqs. (1-2), 
we used the standard quenching factor of 0.77 for the $\sigma \tau$ 
operator.\cite{[CPZ90]}
We used up to 250 intermediate $1^+$ $^{48}$Sc states in the sum. 
They exhaust nearly the entire 
B(GT) sum rules for the transitions from $^{48}$Ti and $^{48}$Ca: 1.59 out of the
exact 1.6 for Ti and 22 out the exact 24 for Ca.

We also tested the  validity of the quenching factor of 0.77 by comparing the
beta decay probabilities for the $^{48}$ Sc$(6^+)\rightarrow ^{48}$Ti$(6^+)$ transitions
with the experimental data.\cite{nds-a48} The results are presented in Table I and confirm that 
this value reasonably describes the B(GT) quenching for this mass region.

\begin{table}
\caption{Theoretical end experimental $logft$ for the 
$^{48}$Sc$\rightarrow ^{48}$Ti beta decay transitions.}
\begin{tabular}{|c|c|c|}
\hline
\hline
$E_x(6^{+}\ ^{48}$Ti) &   $log(ft)_{exp}$ &  $log(ft)_{GXPF1A}$ \\
\hline
3.333       &    5.247       &    5.532\\
 3.508     &      6.083     &      6.010\\
\hline
\hline
\end{tabular}
\vspace*{0.5cm}
\end{table}

The running NME, $M^{2\nu}_{GT}$ of Eqs. (1-2), as a function
of the excitation energy of the $1^{+}$ states in $^{48}$Sc are presented
in Figs. 1 and 2, respectively. The convergence trends are similar to the ones
found in Refs. \cite{[ZBR90], [npajp]}, and it is also supported by the nearly exhausted
sum rules. It is also clear that the phases of the intermediate 
states in the double sum play an essential role: the contribution of the
intermediate states is not coherent and the sum is not continuously 
increasing.\cite{[recentexp]} This is illustrated more clearly
in Fig. 3 where the coherent sum (signed amplitudes)
is compared to the incoherent sum (absolute GT amplitudes).
Fig. 4 presents the relevant
B(GT) transition probabilities. We note that most of the positive
contribution to the double-beta matrix element to the 
$^{48}$Ti ground state 
shown in Fig. 1 comes
from two out of the five intermediate $1^{+}$ states below 5 MeV 
excitation in $^{48}$Sc. For the product 
$[\langle 0_f\vert\vert \sigma\tau^-\vert\vert 1^+_{k} \rangle]
[\langle 1^+_{k}\vert\vert \sigma\tau^- \vert\vert 0_i\rangle]$
the lowest 1$^+$ (at 2.5 MeV) gives [0.185][1.15] = 0.122
and the third 1$^+$ (at 3.8 MeV) gives [0.42][0.35] = 0.147.

Recent experiments have attempted to extract B(GT) values from
$^{48}$Ca($^3$He,t)$^{48}$Sc and 
$^{48}$Ti(d,$^2$He)$^{48}$Sc reaction
cross sections
(see e.g. Fig. 4 of a recent review, Ref. \cite{frekers}).
The results obtained for the
lowest strong 1$^+$ observed in $^{48}$Ca($^3$He,t) at 2.5 MeV
[B(GT)$^{1/2}$ for the above product] is 
$\vert 0.12(3)\vert \times \vert 0.95(5)| = \vert 0.11(3)\vert$, 
in good agreement with
theory given the uncertainties that exist in extracting
B(GT) from charge-exchange cross sections \cite{zegers} (the state
at 2.2 MeV in $^{48}$Ti(d,$^2$He) associated with 1$^+$
does not have a correspondence in the theory -  it is
near a state previously assigned 3$^+$ in the literature
and its $J^{\pi}$ value should be confirmed.) The double-beta strength 
associated with the theoretical state at 3.8 MeV appears to
be spread over several states near 3 MeV in experiment.

Using the results from Figs. 1-2 one gets the following converged results
for the the $2\nu \beta\beta$ matrix elements: 
\begin{itemize}
\item
$\mid M(0^{+}\rightarrow 0^{+})\mid$ = 0.0539 MeV$^{-1}$  for GXPF1A, and
     $\mid M(0^{+}\rightarrow 0^{+})\mid$ = 0.0635 MeV$^{-1}$  for GXPF1;
\item
 $\mid M(0^{+}\rightarrow 2^{+})\mid$ = 0.0122 MeV$^{-3}$  for GXPF1A, and
    $\mid M(0^{+}\rightarrow 2^{+})\mid$ = 0.0129 MeV$^{-3}$  for GXPF1.
\end{itemize}

Using these matrix elements and the phase factors of Ref. \cite{[SC98]} in
Eqs. (3) one gets for the $2\nu \beta\beta$ decay half-lives:
\begin{itemize}
\item
 $T_{1/2}(0^{+}\rightarrow 0^{+})$ = $3.3\times 10^{19}$ $yr$   for GXPF1A, and
     $T_{1/2}(0^{+}\rightarrow 0^{+})$ = $2.4\times 10^{19}$ $yr$  for GXPF1;
\item
 $T_{1/2}(0^{+}\rightarrow 2^{+})$ = $8.5\times 10^{23}$ $yr$ for  GXPF1A, and
    $T_{1/2}(0^{+}\rightarrow 2^{+})$ = $7.5\times 10^{23}$ $yr$   GXPF1.
\end{itemize}

Our value for  the  $T_{1/2}(0^{+}\rightarrow 0^{+})$, corresponding to the 
$\mid M(0^{+}\rightarrow 0^{+})\mid$ = 0.0539 MeV$^{-1}$ NME 
calculated with GXPF1A interaction, is within the 
experimental range \cite{[BAL96]}: 
$(4.3^{-1.3}_{+3.3})\times 10^{19}$ $yr$.
The calculations performed with the GXPF1 interaction seem to give a larger value 
for the NME that leads to a half-life value which is just below the 
present experimental range. 
Comparing our results to the previous similar ones of Refs. \cite{[ZBR90]} 
and \cite {[CPZ90]} we note that
the calculations of Zhao, Brown and Richter are performed in a restricted $pf$ 
model space, and they found a NME of
0.07 MeV$^{-1}$ as their best  value.
Furthermore, using a phase space factor slightly different from ours, 
they obtain $T_{1/2}(0^{+}\rightarrow 0^{+})$ = $1.9\times 10^{19}$ $yr$. 
This half-life is about half of our value and is significantly below
the experimental range.    

Caurier, Poves and Zuker found $\mid M(0^{+}\rightarrow 0^{+})\mid$ = 0.0402 MeV$^{-1}$ 
in their work \cite{[CPZ90]}. 
We repeated their calculations as described 
in Ref.  \cite{[CPZ90]}, and obtained 0.047 MeV$^{-1}$ for the same NME, 
value that is in agreement 
with the NME reported by Nowacki in Ref. \cite{[NOW05]}, and which differs by
about 13\% from ours. This value is also within the present experimental range.   

For the g.s. to the $2^+_1$ transition we obtained a NME value, which 
is about half
the numerical value reported by
Caurier, Poves and Zuker in Ref. \cite{[CPZ90]}. 
However, Ref. \cite{[CPZ90]} used the same Eq. (1) 
for the $\mid M(0^{+}\rightarrow 2^{+})\mid$ NME, 
instead of our Eq. (2), which is recommended
in the literature \cite{[SC98],[STO]}. 
Based on Eq. (1), Ref. \cite{[CPZ90]} suggests that
the $0^{+}\rightarrow 2^{+}$ decay rate is about 3\% of the
$0^{+}\rightarrow 0^{+}$ decay rate.
Our value obtained for the $2\nu\beta\beta$ 
half-life corresponding to the transition 
to the $2^+_1$ excited state of $^{48}$Ti from Eq. (2)
is about four orders of magnitude 
larger than that for the g.s. 
to g.s transition.

  
\begin{figure*}
\includegraphics[width=7cm,angle=0]{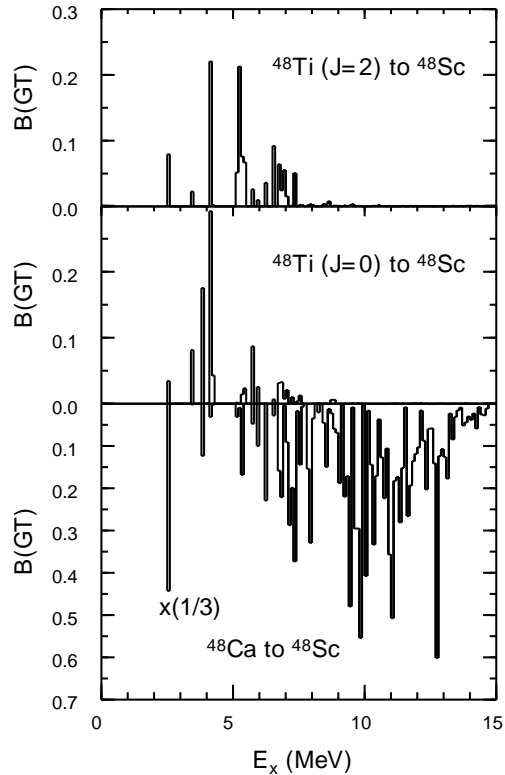}
\caption{\label{f4} The relevant B(GT) probabilities for transitions to the  $1^{+}$ states in
$^{48}$Sc.} 
\end{figure*}

In conclusion we calculated the NME and half-lives for $2\nu\beta\beta$ decay of $^{48}$Ca  
within a SM approach in the full $pf$ model space. We calculated both the g.s. to g.s. 
and g.s. to $2^+_1$ excited state transitions. We use for the first 
time\cite{poves06} 
in such calculations the two versions of GXPF1 two-body interaction, which were recently 
proposed and successfully used to reproduce the 
spectroscopic properties of many nuclei in the 
nuclear mass range A=47-66. Our results are based on
 250 $1^+$ intermediate states in 
$^{48}$Sc nucleus which are enough to exhaust 
almost the entire B(GT) sum rules for the 
transitions from $^{48}$Ti and $^{48}$Ca. 
We also checked the validity of 0.77 quenching factor for the Gamow-Teller 
operator used in the SM calculations 
by comparing the calculated beta transitions 
$^{48}$Sc$\rightarrow ^{48}$Ti with the experimental ones.
The best values, we propose for NME are 
$\mid M(0^{+}\rightarrow 0^{+})\mid$ = 0.0539 MeV$^{-1}$
and $\mid M(0^{+}\rightarrow 2^{+})\mid$ = 0.0122 MeV$^{-3}$,
which were obtained using the GXPF1A interaction. 
They correspond to  $T_{1/2}(0^{+}\rightarrow 0^{+})$ = $3.3\times 10^{19}$ $yr$ and  
$T_{1/2}(0^{+}\rightarrow 2^{+})$ = $8.5\times 10^{23}$ $yr$, respectively. 
Future experiments on $\beta\beta$ decay of $^{48}$Ca,
CANDLES\cite{candles} and CARVEL\cite{carvel}, may reach the required sensitivity of
measuring such transitions and our results could be useful for the planning of
these experiments.
   
\vspace{0.8cm}
\noindent
MH and BAB acknowledge support from the
NSF grant PHY-0555366. SS acknowledge travel support
from the NSF grant INT-0070789.

\end{document}